\documentclass[aps,prl,twocolumn,showpacs,groupedaddress,amsmath,amssymb]{revtex4}
\usepackage{graphicx}
\usepackage{color}
\usepackage{bm}
\usepackage{latexsym}

\begin{document}

\title{Unidirectional Lasing Emerging from Frozen Light in Non-Reciprocal Cavities}
\author{H. Ramezani$^{1}$, S. Kalish$^{1}$, I. Vitebskiy$^{2}$, T. Kottos$^{1}$}
\affiliation{$^1$Department of Physics, Wesleyan University, Middletown, CT-06459, USA}
\affiliation{$^2$Air Force Research Laboratory, Sensors Directorate, Wright Patterson
AFB, OH 45433 USA}
\date{\today }

\begin{abstract}
We introduce a class of unidirectional lasing modes associated with the frozen mode regime 
of non-reciprocal slow-wave structures. Such asymmetric modes can only exist in cavities 
with broken time-reversal and space inversion symmetries. Their lasing frequency coincides with 
a spectral stationary inflection point of the underlying passive structure and is virtually independent 
of its size. These unidirectional lasers can be indispensable components of photonic integrated circuitry.
\end{abstract}

\maketitle

In the lasing process, a cavity with gain produces outgoing optical fields with a definite frequency 
and phase relationship, without being illuminated by coherent incoming fields at that frequency 
\cite{H86}. Instead, the laser is coupled to an energy source (the pump) that inverts the electron
population of the gain medium, causing the onset of coherent radiation at a threshold value of the 
pump. In most cases the first lasing mode can be associated with a passive cavity mode. The latter 
is determined by the geometry and the electromagnetic constitutive parameters $\epsilon-\mu$ 
of the passive cavity. Above the first lasing threshold, lasers have to be treated as nonlinear systems 
\cite{stone}, but up to the first threshold they satisfy the linear Maxwell equations with a negative imaginary 
part of the refractive index, generated by the population inversion due to the pump \cite{H86}. This 
simplification allows for a linear treatment of threshold modes within a scattering matrix formalism \cite{F00}.

In this Letter, using scattering formalism, we introduce a class of unidirectional 
lasing modes emerging from frequencies associated to spectrally asymmetric stationary inflection points 
of the underlying passive photonic structure. A distinctive characteristic of these modes is that, in 
contrast to traditional Fabry-Perot (FP) resonances, they are virtually independent of the size and geometry 
of the confined photonic structure \cite{FV11}. Utilizing these modes we propose to create a Mirrorless 
Unidirectional Laser (MUL) which emits the outgoing optical field into a single direction. Incorporating a 
MUL in an optical ring resonator can result in various functionalities. Potential applications include 
optical ring gyroscopes in which a beat frequency between two oppositely directed unidirectional 
ring diode lasers is detected to measure the rotation rate, optical logic elements in which the direction of
lasing in the ring is the logic state of the device, and optical signal routing elements for photonic 
integrated circuits where the signals are routed around a ring cavity towards a specific output coupler.

\begin{figure}[tbp]
\begin{center}
\includegraphics[scale=0.4]{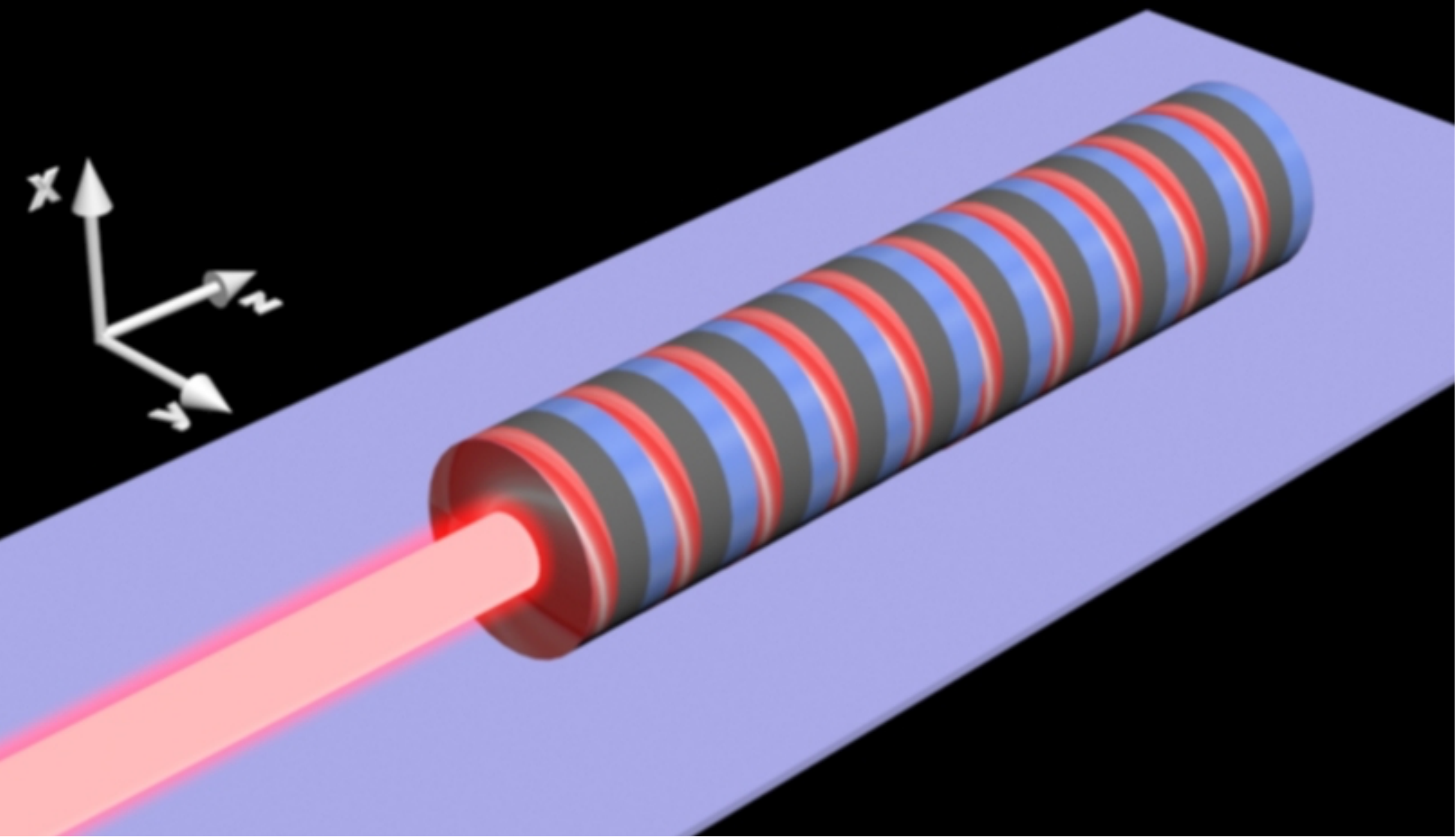}
\end{center}
\caption{Schematic of a  Mirrorless 
Unidirectional Laser structure. The basic unit includes three layers-- a gyrotropic magnetic element 
layer (gray) sandwiched between two misaligned anisotropic layers (red and blue). The misalignment angle 
$\protect\phi _{1}-\protect\phi _{2}$ must be different from $0$ and $\protect \pi /2$.}
\label{fig1}
\end{figure}

The concept of frozen modes and electromagnetic unidirectionality first emerged within the context of spectral 
asymmetry of nonreciprocal periodic structures \cite{F65,D70,VEBL97}. In this regard it was recognized that
magnetic photonic crystals satisfying certain symmetry conditions \cite{VEBL97,FV01} can develop a strong spectral 
asymmetry $\omega ({\vec{k}})\neq \omega (-{\vec{k}})$. An example case of such periodic arrangement is shown 
in Fig. \ref{fig1}. The basic unit consists of three components: a central magnetic layer \textquotedblleft sandwiched" 
between two misaligned anisotropic layers. The magnetic layer (gray layer in Fig. \ref{fig1}) induces magnetic 
non-reciprocity which is associated with the breaking of time reversal symmetry due to a static 
magnetic field or spontaneous magnetization. However, breaking time-reversal symmetry is not sufficient to obtain 
spectral asymmetry. The symmetry analysis \cite{VEBL97,FV01} shows that the absence of space inversion is also 
required. This is achieved with the use of two birefringent layers (blue and red layers in Fig. \ref{fig1}).

The constitutive tensors $\hat\epsilon_{1,2}-\mu$ of the nonmagnetic layers
are assumed to be 
\begin{equation}  
\label{permittivity}
\hat {\mathbf{\epsilon}}_{1,2}= \begin{bmatrix}
\epsilon_A+\delta\cos(2\phi_{1,2})&\delta \sin(2\phi_{1,2})&0\\ \delta
\sin(2\phi_{1,2})&\epsilon_A-\delta\cos(2\phi_{1,2}) &0\\ 0&0&\epsilon_{zz}
\end{bmatrix};  \hat\mu =\hat{\mathbf{1}}
\end{equation}
where $\delta$ describes the magnitude of in-plane anisotropy and the angle 
$\phi_{1,2}$ defines the orientation of the principle axes in the $xy-$plane
for each of the two nonmagnetic layers. The corresponding constitutive
parameters for the magnetic layer are 
\begin{equation}  
\label{permiability}
\hat{\epsilon}= 
\begin{bmatrix} 
\epsilon_F&i\alpha&0\\
-i\alpha&\epsilon_F&0\\ 0&0&\epsilon_{zz} 
\end{bmatrix}; \hat{\mu}= 
\begin{bmatrix} 
\mu_{xx}&i\beta&0\\ -i\beta&\mu_{xx}&0\\ 0&0&\mu_{zz}
\end{bmatrix};
\end{equation}
The gyrotropic parameters $\alpha,\beta$ are responsible for the magnetic Faraday rotation \cite{LL84}. In the simulations 
below we use the same set of physical parameters as in \cite{FV01}. Specifically: $\epsilon_A = 43.85, \delta = 42.64,  \phi_1 = \pi/4,  
\phi_2 = 0, \epsilon_F = 30.525,  \alpha = 0.625, \beta = 1.24317$, and $\mu_{xx} = 1.29969$. The frequency is measured in 
units of $\Omega = c L$, where $c$ is the speed of light in vacuum and $L=L_A$ is the thickness of the dielectric layers. The width 
of the ferromagnetic layer is $L_F = 0.45 L$. Without loss of generality we assume that $L= 1$. 

The dispersion relation $\omega ({\vec{k}})$ for the infinite periodic stack of Fig. \ref{fig1} can be calculated numerically using a 
standard transfer matrix approach \cite{FV01,RLKKKV12}. We find that $\omega ({\vec{k}})$ displays asymmetry with respect to 
the Bloch wave vector ${\vec{k}}$. For a given structural geometry, the degree of the spectral asymmetry depends on the magnitude 
of nonreciprocal circular birefringence of the magnetic layer and linear birefringence of the misaligned dielectric layers. If either 
of the two birefringences is too small, or too large, the spectral asymmetry becomes small. The choice of the
numerical parameters allows 
as to clearly demonstrate the effect of unidirectional lasing using the simplest example of a periodic layered structure shown in 
Fig. \ref{fig1}. At optical frequencies, very similar results can be achieved by turning the open cavity of Fig. \ref{fig1} into a 
ring-like structure \cite{note1}.

\begin{figure}[tbp]
\begin{center}
\includegraphics[scale=0.35]{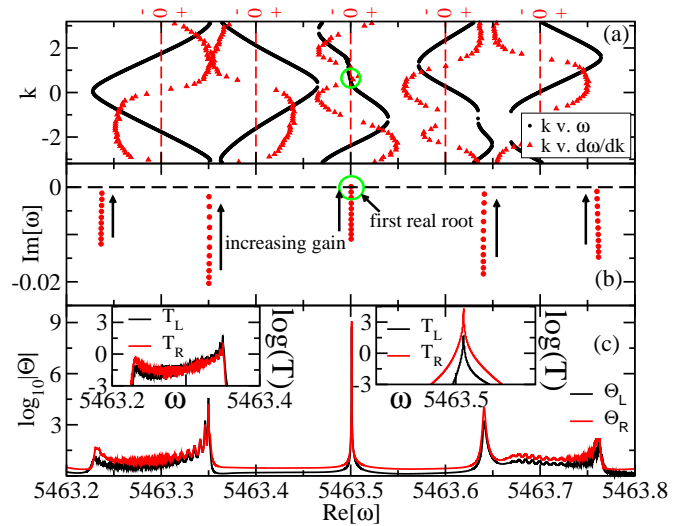}
\end{center}
\caption{(a) Asymmetric dispersion relation $\omega(k)$ (black filled circles) of the passive magnetic photonic structure shown in Fig. \ref{fig1} . 
Overlaid is the group velocity $\frac{d\protect\omega }{dk}$ of the propagating modes (red triangles). The SIP is indicated with a circle. 
(b) Motion of exact S-matrix poles in the complex k-plane as the gain inside the photonic structure increases from zero. The arrows 
indicate the direction of increasing gain. The green circle marks the first lasing threshold. (c) The response function $\Theta_{L,R}(\omega)
=T_{L,R}+R_{L,R}$ vs. $\protect\omega $ for a grating consisting of $20$ basic units. Left inset: $T_L$ left (black) and $T_R$ 
right (red) transmittance for a frequency domain where Fabry-Perot resonances are present. Right inset: $T_{L,R}$ around the
SIP frequency. Notice the high level of asymmetry near the SIP where $T_R\gg T_L$ (more than two order of magnitude).
}
\label{fig2}
\end{figure}

The property of spectral asymmetry has various physical consequences, one of
which is the possibility of unidirectional wave propagation \cite{FV01}. Let us
consider a transverse monochromatic wave propagating along the symmetry
direction ${\hat{z}}$ of the gyrotropic photonic crystal. The Bloch wave
vector ${\vec{k}}=k{\hat{z}}$, as well as the group velocity ${\vec{v}}({\vec{k}})\equiv \partial 
\omega ({\vec{k}})/\partial {\vec{k}}$ are parallel to $z$. In Fig. \ref{fig2}a we see that one of the spectral branches $\omega
(k)$ develops a stationary inflection point (SIP) for which 
\begin{equation}
{\frac{\partial \omega }{\partial k}}|_{k=k_{0}}=0;\quad {\frac{\partial
^{2}\omega }{\partial k^{2}}}|_{k=k_{0}}=0;\quad {\frac{\partial ^{3}\omega 
}{\partial k^{3}}}|_{k=k_{0}}\neq 0  \label{sip}
\end{equation}%
An example of a SIP occurring at $\omega _{0}\approx 5463.5$ is marked in Fig. \ref{fig2}a with a circle. At 
this frequency there are two propagating Bloch waves: one with $k_{0}\approx 0.613$ and the other with $k_{1}\approx 
-2.452$. Obviously, only one of the two waves can transfer electromagnetic energy - the one with $k=k_{1}$ and corresponding 
group velocity pointing in the positive ${\hat{z}}$ direction i.e. ${\vec{v}}(k_{1})>0$. The Bloch eigenmode with $k=k_{0}$ 
has zero group velocity ${\vec{v}}(k_{0})=0$ and therefore does not transfer energy. The latter propagating mode is associated 
with a stationary inflection point Eq. (\ref{sip}) and referred to as the ``frozen'' mode. Thus a photonic crystal with the dispersion 
relation similar to that in Fig. \ref{fig2}a displays the property of electromagnetic unidirectionality at $\omega =\omega _{0}$. Such
a remarkable effect is an extreme manifestation of the spectral asymmetry \cite{FV01}.

Next we turn to the analysis of the open photonic structure and the emergence of
unidirectional lasing modes. It can be rigorously shown within semiclassical
laser theory that the first lasing mode in any cavity is an eigenvector of
the electromagnetic scattering matrix (${\mathbf{S}}$ -matrix) with an
infinite eigenvalue; i.e., lasing occurs when a pole of the ${\mathbf{S}}$-matrix is pulled ``up'' to the real axis by including gain as a negative
imaginary part of the refractive index \cite{F00}. We therefore proceed to
evaluate the scattering matrix associated with the open photonic structure of Fig. \ref{fig1}.

The electric $\vec{\mathbf{E}}(\vec{r})$ and magnetic $\vec{\mathbf{H}}(\vec{r})$ fields are designated by the time-harmonic Maxwell equations: 
\begin{equation}
\nabla \times \vec{\mathbf{E}}(\vec{r})=i\frac{\omega}{c}\hat{\mu}\vec{\mathbf{H}}(\vec{r}), \; 
\nabla \times \vec{\mathbf{H}}(\vec{r}) =-i\frac{\omega}{c}\hat{\varepsilon}\vec{\mathbf{E}}(\vec{r})
\end{equation}
with the solution 
\begin{equation}  \label{eq:1}
\vec{\mathbf{E}}(\vec{r})=e^{i(k_xx+k_yy)}\vec{E}(z), \; \vec{\mathbf{H}}(\vec{r})=e^{i(k_xx+k_yy)}\vec{H}(z)
\end{equation}
where $\omega({\vec k})$ is the frequency. An ``external region'' encompassing
the resonator, extends for $|z|>L/2$. We assume that the permittivity and
permeability parameters take constant values $\hat {\mathbf{\epsilon}}=\epsilon_0\times \hat{\mathbf{1}}; 
\hat {\mathbf{\mu}}= \mu_0\times \hat{\mathbf{1}}$ where $\hat{\mathbf{1}}$ is the $3\times 3$ identity matrix.

Assuming normal propagation, i.e. $k_x=k_y=0$, the solutions of Eq. \eqref{eq:1} for $\vec{E}(z)$ in the left $(l)$ and right $(r)$ side of the
scattering region, are written in terms of the forward and backward traveling waves: 
\begin{equation}
\vec{E}^{l,r}=\mathbf{A}^{l,r}e^{ikz}+\mathbf{B}^{l,r}e^{-ikz}
\end{equation}
where 
\begin{equation}
\mathbf{A}^{l,r}=[A^{l,r}_x \;\; A^{l,r}_y]^T;\quad \mathbf{B}^{l,r}=[B^{l,r}_x \;\; B^{l,r}_y]^T
\end{equation}
The corresponding magnetic field $\vec{H}(z)$ is $\vec{H}(z)=\frac{1}{c}\hat{z}\times \vec{E}(z)$, where $\hat z$ is the unit vector in the $z$-direction.

The relation between the electric field on the left and right sides of the
scattering region is described by the $4\times$4 transfer matrix $\mathbf{M}$: 
\begin{equation}
\begin{bmatrix}{ \textbf{A}}^r \\ \textbf{B}^r 
\end{bmatrix} =\mathbf{M} 
\begin{bmatrix} \textbf{A}^l \\ \textbf{B}^l 
\end{bmatrix};\quad \mathbf{M}= 
\begin{bmatrix} \textbf{M}_{11} & \textbf{M}_{12} \\ \textbf{M}_{21} &
\textbf{M}_{22} \end{bmatrix}
\end{equation}

The transmission and reflection coefficients can be then expressed in terms
of the transfer matrix elements as: 
\begin{equation}
\begin{array}{cc}
{\mathbf{r}}^l=-{\mathbf{M}}_{22}^{-1}{\mathbf{M}}_{21}, & \mathbf{r}^r=\mathbf{M}_{12}\mathbf{M}_{22}^{-1}  \\
\mathbf{t}^l=\mathbf{M}_{11}-\mathbf{M}_{12}{\mathbf{M}_{22}}^{-1}\mathbf{M}_{21}, & \mathbf{t}^r=\mathbf{M}_{22}^{-1}.
\end{array}
\label{eq:2}
\end{equation}
From Eqs. \eqref{eq:2}, we construct the scattering matrix $\mathbf{S}$ 
\begin{equation}
\begin{bmatrix} 
\textbf{B}^l \\ \textbf{A}^r 
\end{bmatrix} =\mathbf{S} 
\begin{bmatrix} \textbf{A}^l \\ \textbf{B}^r 
\end{bmatrix};\quad \mathbf{S}
\equiv \begin{bmatrix} \textbf{r}^l & \textbf{t}^r \\ \textbf{t}^l &
\textbf{r}^r \end{bmatrix}  \label{eq:3}
\end{equation}
which, below the laser threshold, connects the outgoing wave amplitudes to their incoming counterparts. For lossless media 
the permittivity $\epsilon=\epsilon^{\prime }+i\epsilon^{\prime \prime }$ is strictly real $\epsilon^{\prime \prime }=0$. 
Addition of gain in the system results in a complex permittivity $\epsilon=\epsilon^{\prime }+i\epsilon^{\prime \prime }$
with $\epsilon^{\prime \prime }<0$. Without loss of generality we further assume that the gain is introduced at the 
dielectric layers with $\phi_2=0$. We have checked that other periodic arrangements of the gain along the cavity give the same 
qualitative results.

Following Refs. \cite{F00,LGYCSR11} we analytically continue the scattering matrix $\mathbf{S}$ to the complex$-k$ plane. 
The scattering resonances are then defined by the following (outgoing) boundary conditions, i.e. $(\mathbf{B}^l, 
\mathbf{A}^r)^T\neq 0$ while $(\mathbf{A}^l,\mathbf{B}^r)^T=0$, and are associated with the poles of the $\mathbf{S}$ 
matrix. It follows from Eqs. (\ref{eq:2},\ref{eq:3}) that the poles of the ${\mathbf{S}}$-matrix can be identified with the 
complex zeros of the secular equation $\det (\mathbf{M}_{22})=0$.

In a passive structure, where $\epsilon =\epsilon ^{\prime }$ is real, the complex poles of the $\mathbf{S}$ matrix $k_{p}=
k_{R}-ik_{I}$ are located in the lower half part of the complex plane due to causality. The real part $k_R\equiv{\cal R}e(k_p)$ is associated 
with the resonant frequencies while the imaginary part $k_I\equiv{\cal I}m(k_{p})$ describes the fact that the cavity is open \cite{LGYCSR11}.

In Fig. \ref{fig2}b we report the motion of the $\mathbf{S}$-matrix poles in the complex $k$-plane. Introducing gain $\epsilon^{\prime \prime }$ 
to the system leads to an (almost) vertical movement of the poles towards the real axis. This indicates that the lasing frequency ${\cal R}e(k_p)$ 
of the system is almost equal to the associated mode of the passive structure. At the critical value $\epsilon_{\mathrm{Th}}^{\prime\prime }$ 
for which the first one of the poles (marked with a circle in Fig. \ref{fig2}b) crosses the real axis the system reaches the lasing
threshold. We have further confirmed the lasing action at $\epsilon_{\mathrm{Th}}^{\prime \prime }$ by evaluating 
in Fig. \ref{fig2}c an overall response function, defined as the total intensity of outgoing (reflected or transmitted) waves
for either a left or right (single-port) injected wave; that is $\Theta_{L,R}(\omega)=T_{L,R}+R_{L,R}$ where $T_{L/R}$ 
and $R_{L/R}$ are the respective left and right transmittances and reflectances averaged over polarization. For a lossless passive 
medium, one always has $\Theta(\omega)=1$ due to power conservation, whereas $\Theta(\omega)>1$ indicates that 
an overall amplification has been realized. Near the lasing frequency $\Theta(\omega)$ takes large values, diverging as the
lasing threshold is attained. Furthermore in the insets of Fig. \ref{fig2}c we report the left and right transmittances in the regime
of regular FP resonances (left inset) and at a SIP-related frozen mode (right inset). While for FP resonances $T_L$ and 
$T_R$ exhibit a moderate asymmetry, for a SIP-related mode the asymmetry between them increases dramatically ($T_R\gg T_L$
by more than two orders of magnitude) indicating strongly asymmetric transport.

\begin{figure}[tbp]
\begin{center}
\includegraphics[scale=0.325]{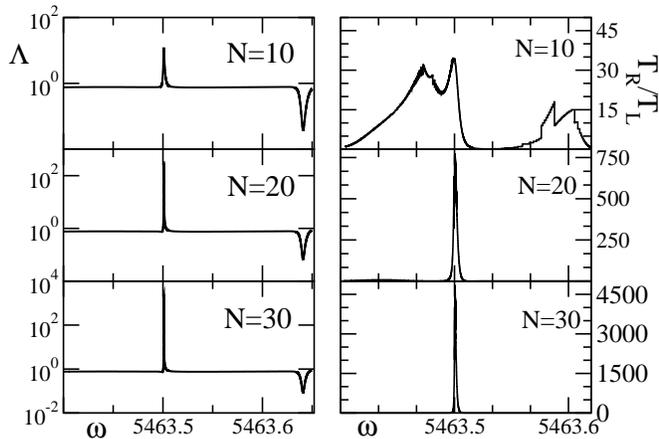}
\end{center}
\caption{The ratio $\Lambda$ (left column) and the associated transmittance asymmetry $T_A=T_R/T_L$ (right column) for the photonic 
structure of the Fig. \protect\ref{fig1}, for systems of (a) 10; (b) 20 and (c) 30-basic units. Notice that $T_A$ increases near the SIP-related 
mode (while it subside at regular FP resonances where $T_A\approx 1$--see also insets of Fig. \ref{fig2}c)}
\label{fig3}
\end{figure}

A comparison between Fig. \ref{fig2}a and Figs. \ref{fig2}b,c indicates that the lasing threshold frequency is very 
close to $\omega_{0}$ associated with the SIP. While, at this frequency, a right-moving propagating wave (associated to 
$k_1$ and having large group velocity $v(k_1)>0$) releases most of its energy outside the photonic structure, the mode 
associated to $k\approx k_0$ has extremely small group velocity and allows for a long residence time of the photons 
inside the structure. The interaction of these photons with the gain medium results in strong amplification which 
in turn leads to a lasing action. Once the lasing threshold is reached, the outgoing lasing beam is emitted predominantly 
from the left side of our structure (opposite side from $v(k_1)>0$), therefore producing unidirectional lasing. 

We have confirmed the asymmetry in overall power amplification by analyzing the ratio $\Lambda\equiv\frac{T_R+R_L}{T_L+R_R}$.
The latter is shown in the right column of Fig. \ref{fig3}. Specifically, large values of $\Lambda$ indicate asymmetric power amplification 
towards the left while smaller than unity $\Lambda$ indicates power amplification towards the right. However the actual indication
of  the {\it non-reciprocal} nature of our photonic structure is provided by an analysis of the transmittance asymmetry $T_A=
T_R/T_L$. This asymmetry exists only in the case that non-reciprocity and gain are simultaneously present. 

The magnitude of $T_A$ will be strongly affected by the existence of the SIP. Specifically we expect that the unidirectional 
amplification near the SIP can be enhanced further by increasing the size of the periodic structure. In this case, the residence time 
of photons associated with the left moving (slow velocity) mode becomes even longer thus resulting in stronger transmission 
asymmetries. A scaling analysis of $T_A$ for an increasing number of periods of the grating (see Fig.~\ref{fig3} right panels) 
indicates that as the system becomes larger the asymmetry becomes increasingly peaked around the unidirectional lasing mode 
(associated to the SIP), while other picks (associated to standard FP resonances) remain unchanged or even subside. We observe 
that the unidirectional  lasing frequency emerging from of the SIP mode  remains fixed and it is insensitive to the size of the 
periodic structure.

\textit{Conclusions - } We have introduced a qualitatively new mechanism for unidirectional lasing action which relies on the 
co-existence of highly non-reciprocal SIP-related frozen modes and gain. The transmittance asymmetry of these modes 
increases significantly with the size of the structure.  As opposed to a conventional lasing cavity utilizing asymmetrically 
placed reflecting mirrors, a nonreciprocal magneto-photonic structure can produces unidirectionality even at the first (linear) 
lasing threshold. This unique feature is not available in reciprocal active cavities. Besides, a unidirectional magneto-photonic 
structure does not need mirrors or any other reflectors in order to trap light and reach the lasing threshold.

The nonreciprocal layered structure in Fig. \ref{fig1} is just a toy model of a periodic array capable of developing a spectral SIP. 
It was rather used here in order to demonstrate the proof of principle for the MUL. A future challenging research direction is to propose 
realistic structures that allow for the co-existence of gain and nonreciprocal SIP. While at infrared and optical wavelengths, the gain 
is not a problem, the creation of a well-defined nonreciprocal SIP is not a simple matter and is the subject of a whole new area of 
research. In this respect, a promising way towards building a unidirectional laser could be to add magneto-optical and gain components 
to a periodic optical waveguide displaying a SIP. Examples of such waveguides can be found, for instance, in \cite{gutman}, and 
references therein. This possibility is currently under investigation.

\textit{Acknowledgement - } This work was sponsored by the Air Force Research
Laboratory (AFRL/RYD) through the AMMTIAC contract with Alion Science and
Technology, and by the Air Force Office of Scientific Research, LRIR
09RY04COR and FA 9550-10-1-0433. Valuable comments from Dr. T. Nelson are greatly appreciated.


\begin{thebibliography}{99}
\bibitem{note1} Due to the inherent weakness of Faraday rotation at optical frequencies, a SIP on the $k-\omega$ diagram will always be located 
in close proximity to a photonic band edge. As a consequence, the frozen mode regime in open photonic cavity of Fig. \ref{fig1} will be overwhelmed 
with the much more powerful and nearly perfectly symmetric, giant FP resonance \cite{FV11}, whose intensity scales as $N^4$ ($N$ is the
number of layers). A way to eliminate this and other transmission resonances and preserve the phenomenon of electromagnetic unidirectionality is to 
use a closed ring-like structure, instead of the open one shown in Fig. \ref{fig1}.
\end{thebibliography}
\end{document}